# Measurement of nucleon time-like form factors at DAFNE-2

P. Rossi

Laboratori Nazionali di Frascati - INFN, Via Enrico Fermi 40, 00044 Frascati, Italy

*The study of the electromagnetic form factors of the nucleon in both the space-like and time-like domains plays a key role for the understanding of the internal structure and dynamics of this basic building block of the matter. They are also necessary for the interpretation of many other measurements in reactions involving nucleons. In spite of being under investigation by more than fifty years, form factors are far from being fully exploited and more efforts devoted to their experimental determination are needed. Here, we discuss the possibility of a complete measurement of the nucleon form factors, both moduli and phases, in the time-like region at Frascati with the DA$\Phi$NE electron-positron storage ring upgraded in energy.*

## 1. INTRODUCTION

Reproducing the structure of the nucleon is one of the defining problems of Quantum Chromodynamics. When proton and neutron, which form the building blocks of nuclei in their foundamental state, were discovered they were thought as being Dirac particles, i.e. they were expected to be point-like without any internal structure. Electron scattering experiments in the 1950's revealed, instead, a charge and magnetization distribution inside the nucleon. Two form factors (FFs), called the elastic electric and magnetic FFs, were introduced to account for these distributions.

FFs are not only important because they contain information on the nucleon ground state and thus constitue a severe test for models of nucleon structure but also because they are necessary for the interpretation of many other measurements in reactions involving nucleons. In particular, elastic FFs are a limiting case of the Generalized Parton Distribution (GPD) functions [1], the universal non-perturbative objects describing hard exclusive processes induced by photons and electrons or positrons, and can be used to constraint GPD models [2].

FFs are defined by the matrix elements of the electromagnetic current $J_\mu(x)$ of the nucleon (as in Fig. 1):

$$\langle N(p') | J^\mu(0) | N(p) \rangle = e\bar{u}(p') \left[ F_1(q^2)\gamma^\mu + \frac{i}{2M} F_2(q^2)\sigma_{\mu\nu}q^\nu \right] u(p) \tag{1}$$

where $e$ is the proton charge, $M$ is the nucleon mass and $q^2 = (p - p')^2$ is the squared momentum transfer in the photon-nucleon vertex. The Dirac ($F_1$) and Pauli ($F_2$) FFs, defined in terms of elements of the Dirac equation, are normalized so that for the proton $F_1^p(0) = 1$ and $F_2^p(0) = \kappa_p = \mu_p - 1$ and for the neutron $F_1^n(0) = 0$ and $F_2^n(0) = \kappa_n = \mu_n$, where $\kappa_p$ and $\kappa_n$ are the anomalous part of proton and neutron magnetic moments, $\mu_p$ and $\mu_n$ respectively. Eq. (1) can be rearranged by defining the Sachs electric ($G_E$) and magnetic ($G_M$) FFs:

$$G_E = F_1 + \tau F_2 \qquad\qquad G_M = F_1 + F_2 \qquad (2)$$

where $\tau = q^2/4M^2$ and their normalization is $G_E^p(0) = 1$, $G_M^p(0) = \mu_p$, $G_E^n(0) = 0$, $G_M^n(0) = \mu_n$.

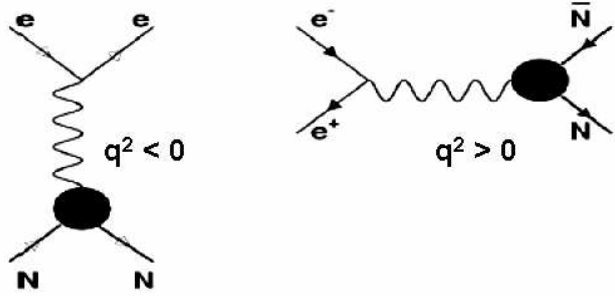

Figure 1. *Diagrams for space-like (left) and time-like (right) electromagnetic nucleon FF measurements.*

Each form factor is assumed to be described by a function which is analytic in the complex $q^2$ plane. Space-like FFs ($q^2 < 0$) are real while time-like FFs ($q^2 > 0$) have a phase and they are connected by dispersion relations [3].

## 2. THE NUCLEON SPACE-LIKE FORM FACTORS

The elastic lepton-nucleon scattering gives access to the space-like FFs. Basically, the FFs are extracted using the Rosenbluth technique [4] which consists by linearly fitting the $\epsilon$-dependence of the cross section at fixed $Q^2$:

$$\frac{d\sigma}{d\Omega} = \left(\frac{d\sigma}{d\Omega}\right)_{Mott} \frac{\tau}{\epsilon(\tau-1)} \left(G_M^2 - \frac{\epsilon}{\tau}G_E^2\right) \qquad (3)$$

where $\epsilon$ is the virtual photon polarization. The measurements for proton [5] (empty triangles in Fig. 2) indicate an approximate FF scaling, i.e. $\mu_p G_E^p/G_M^p \approx 1$ up to $Q^2 \approx 7$ GeV$^2$, though with large uncertainties in $G_E^p$ at the highest $Q^2$ values. In fact, the presence of the factor $1/\tau$ in front of $G_E$ makes difficult its extraction when $Q^2$ becomes large.

More recently, the polarization transfer method [6], which requires longitudinally polarized electron beam and the measurement of the recoil proton polarization has been applied to obtain directly the ratio

$$\mu_p \frac{G_E}{G_M} = -\mu_p \frac{P_T L}{P_L} \frac{E_e + E_{e'}}{2M_p} tan(\theta_e/2) \qquad (4)$$

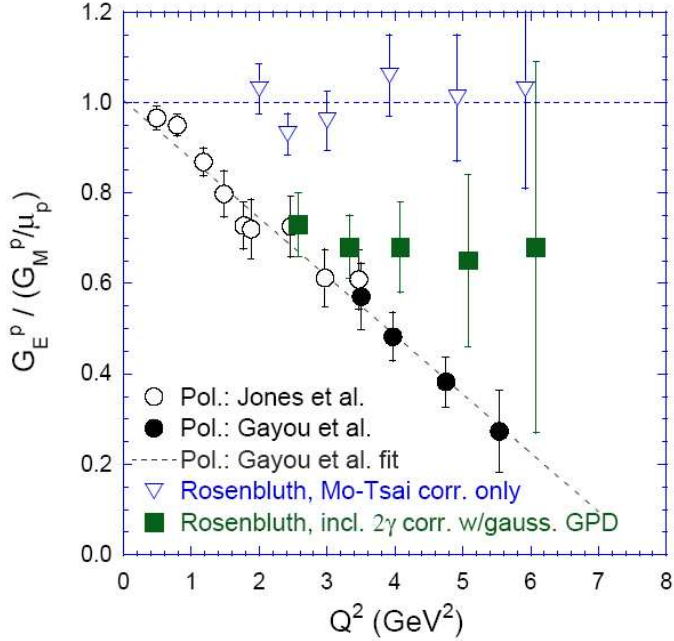

Figure 2. *Comparison of the ratio $\mu_p G_E^p/G_M^p$ of the FFs extracted by Rosenbluth technique (hollow triangles) with recoil polarization measurements (full and empty circles) [8–10]. The dashed line is a linear fit to the polarization transfer data. Full squares are the Rosenbluth measurements corrected for two photon exchange [14].*

where $E_e$ and $E_{e'}$ are the incoming and scattered electron energies, $\theta_e$ is the electron scattering angle, $P_L$ and $P_T$ are the longitudinal and transverse components of the final proton polarization in the hadronic scattering plane. Because in the ratio the two polarizations are simultaneously measured, this technique avoids major systematic uncertainties of the Rosenbluth separation. Moreover, the contribution of the small electric FF is enhanced, because there are no suppression factors with increasing $Q^2$, thus allowing precision measurements also at high $Q^2$.

While the low $Q^2$ measurements at MIT-Bates [7] are consistent with the Rosenbluth data, the JLab measurements [8,9] (empty and full circles in Fig. 2), which extended the data up to $Q^2 \approx 5.6$ GeV$^2$, significantly deviates from the scaling behavior. In fact, they show an almost linear decrease from unity at low $Q^2$ up to $\approx 0.3$ at the highest $Q^2$. The linear fit of these data (dashed curve in Fig. 2) also suggests that the electric FF should cross the zero for $Q^2 \approx 8$ GeV$^2$.

A recent reanalysis of the experimental data obtained with the Rosenbluth technique [11] showed that, the individual measurements are consistent to each other. In addition, the new Rosenbluth measurement performed at JLab [12] for $Q^2$ between 2.6 and 4.1 GeV$^2$ confirmed the FF scaling, making it clear that the source of the discrepancy is not simply an experimental problem.

Recent works [13–15] suggest that the possible origin of these discrepancies is the failure of the one-photon approximation to adequately describe the result of high-precision unpolarized experiments. If two-photon exchange is not negligible, then the Rosenbluth formula will not represent the underlying structure of the proton, but it will be just a parametrization of the *ep* elastic cross section. An estimate done by [14] showed that in the range $Q^2 = 2 - 3$ GeV$^2$ the Rosenbluth method including two-photon exchange correction agrees well with polarization results, and at higher $Q^2$, there is at least partial reconciliation between the two methods (full squares in Fig. 2).

### 3. THE NUCLEON TIME-LIKE FORM FACTORS

The time-like $G_E(s)$ and $G_M(s)$ (here and in the following, we define the total energy squared $s = q^2$) are the analytical continuation of non spin flip $G_E(Q^2)$ and spin flip $G_M(Q^2)$ space-like FFs.

The absolute value of nucleon FFs can be derived from cross section measurements in $e^+e^- \to N\bar{N}$ or $p\bar{p} \to e^+e^-$ experiments. In the one-photon exchange approximation, the center-of-mass total and differential cross section as a function of $s$ are given by [16]:

$$\sigma = \frac{4\alpha^2\pi\beta}{3s}C\left[\mid G_M \mid^2 + \frac{1}{2\tau}\mid G_E \mid^2\right] \tag{5}$$

$$\frac{d\sigma}{d\Omega} = \frac{\alpha^2\beta}{4s}CD = \frac{\alpha^2\beta}{4s}C\left[\mid G_M \mid^2 (1+cos^2\theta) + \frac{1}{\tau}\mid G_E \mid^2 sin^2\theta\right] \tag{6}$$

where $\theta$ is the nucleon scattering angle, $\beta = \sqrt{1 - 4M^2/s}$ is the nucleon velocity, $\tau = s/4M^2$ and $C$ is the Coulomb correction factor [17]. As in the space-like region, at high energy the cross section is dominated by the magnetic term, because of the presence of the factor $1/\tau$ in front of the electric one.

Experimentally, the time-like FFs are poorly known both for proton and neutron.

**Proton**
Proton FF data have been obtained by measuring total cross section in $e^+e^- \to N\bar{N}$ and $p\bar{p} \to e^+e^-$ experiments [18–31] and arbitrarily assuming that the relation $|G_E^p| = |G_M^p|$, theoretically valid at threshold, holds in the whole explored energy region. Actually there are hints that the $|G_E^p|/|G_M^p|$ ratio is not constant with $s$. In fact, despite the large errors, experimental angular distributions [18–21] suggest large variations of $|G_E^p|/|G_M^p|$ above the threshold, thus strongly challenging the relation $|G_E^p| = |G_M^p|$. However, based on this approximation, the magnetic FF has been derived, leaving on the other hand the electric FF basically experimentally unknown.

An unexpected feature of $G_M^p$ is the steep decrease very near to the threshold indicating the presence of a resonance with mass of 1.87 GeV/c which could be identified as a possible $N\bar{N}$ bound state. The origin of the steep behavior at threshold will be settled once a complete set of accurate data on both $G_E^p$ and $G_M^p$, absolute value and relative phase, of the proton and neutron, will be available.

Another somewhat unexpected aspect concerns the general trend of $|G_M^p(q^2)|$: analyticity [32] and pQCD predict the time-like FF is asymptotically close to the space-like value at the same $|Q^2|$ [33]; that is, the time-like proton FF asymptotically becomes real and scales as $1/Q^4$. Experimental data are consistent with the scaling already above $Q^2 \approx 3\text{-}4$

GeV$^2$, but they exceed the space-like value by a factor of 2 [34]. Clearly, the uncertainty about the contribution of the electric FF might strongly affect these conclusions.

Finally, the new BaBar data[31] show a further quite unexpected cross section reaching the asymptotic trend by means of negative steps, at $\sqrt{s} \approx 2.2$ GeV and at $\sqrt{s} \approx 2.9$ GeV.

**Neutron**

The neutron magnetic FF has been measured by the FENICE experiment [18] only from threshold up to $s = 4.4$ GeV$^2$ with low statistics (about 100 $n\bar{n}$ events collected). For the neutron magnetic FF, pQCD predicts the limit $|G_M^n| \approx (q_d/q_u)^2 |G_M^p| = |G_M^p|/4$, that is a cross section $\sigma(e^+e^- \to p\bar{p})$ almost four times bigger than $\sigma(e^+e^- \to n\bar{n})$. On the contrary, despite the large statistical errors, the FENICE data clearly show that above threshold the $n\bar{n}$-production cross section is almost twice as $p\bar{p}$ one, indicating that the neutron magnetic FF should be much bigger than the pQCD extrapolation from proton data. Within the scant FENICE statistics, the neutron electric FF is consistent with zero at threshold. In such a case the neutron magnetic FF should also be zero, and this is consistent with the flat (within the large errors) measured angular distributions.

In conclusione we can say that time-like FFs are poorly known experimentally, the neutron data are very scarce and polarization data, directly connected with the complex phases of FF, are not available at all. Proton FFs themselves, even having been the most extensively studied, rely on approximations that are not justified.

## 4. THE NUCLEON TIME-LIKE FORM FACTORS MEASUREMENT AT DAFNE-2

A physics program to perform the full determination of nucleon form factors in the time sector with the Frascati DAΦNE electron-positron storage ring upgraded in energy (in the following called DAFNE-2) has been proposed by a collaboration of 82 physicists from 25 Institutions of 8 Countries. With this measurement it will be possible to performe the first complete measurement of both moduli and phases of the nucleon FFs from threshold up to 2.4 GeV. This region is particularly suited for discriminating among the different theories all able to well describe the data in the space-like region. This can be clearly seen in Fig. 3 where theoretical predictions for the ratio of electric to magnetic proton FF in the time-like region as a function of the beam energy is reported. The shadow area is the result of the dispersion relation analysis [3] while the other curves are based on QCD models [35].

A preliminary estimate of the expected counting rate and the attainable precision in the FF extraction has been made using the FINUDA detector (which is one of the two detectors hosted in the DAΦNE storage ring). To this aim eq. (6) normalized to the total cross section value and the following assumptions for the two electromagnetic FFs have been used: $|G_E| = |G_M|$ (for proton and neutron) or the value $|G_E|/|G_M|$ given by the dispersion relation analysis of Fig. 3 (for the proton) or $|G_E| = 0$ (for the neutron). From this *theoretical* distribution, the number of events as a function of the nucleon angle has been extraced randomly, obtaining angular distributions like those shown, as an example, in Fig. 4 and 5 for proton and neutron respectively, for three different beam energies and for an integrated luminosity of 100 $pb^{-1}$ per beam energy. For the proton a constant detection efficiency close to 1 for all polar angles, and a geometrical azimuthal acceptance $\epsilon_p \approx 0.85$ has been assumed, while for the neutron a constant detection efficiency of $\epsilon_n = 0.15$ has been used. The angular distributions have been fitted over the FINUDA angular coverage ($\theta = 45^o - 135^o$) with the function

$$f(\theta) = A(1 + cos^2\theta) + B sin^2\theta \tag{7}$$

and $|G_M|$ and $|G_E|$ have been directly extracted from the coefficients $A$ and $B$, respectively. These fits are shown by the full curves in Fig. 4 and 5.

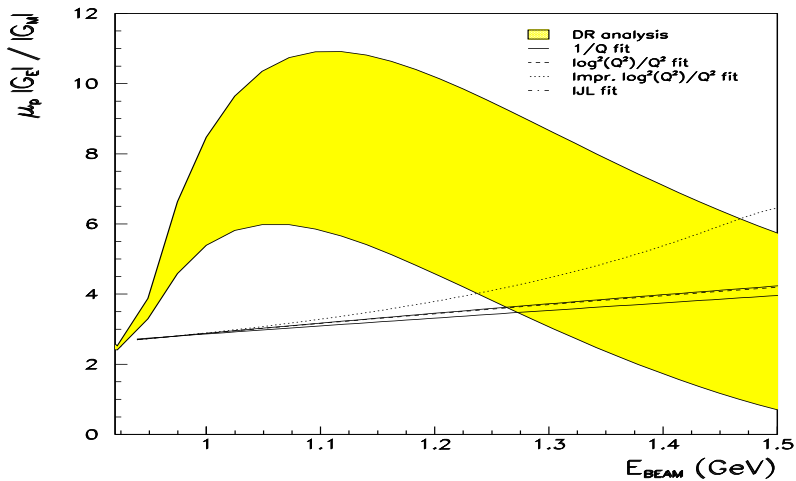

Figure 3. *Theoretical predictions for the ratio of electric to magnetic proton FF in the time-like region as a function of the beam energy. The shadow area is the result of the dispersion relation analysis [3]. The other curves are based on QCD models (see ref. [35] for details).*

In order to properly take into account the correlation between the two coefficients of eq. (7), this procedure has been repeated for several samples of the same distribution. The mean value and the width of these distributions give an estimate of the average value and the error on the FFs extraction. The relative error resulting from this procedure shows that a maximum error of 2% on $|G_M^p|$ and of 4% on $|G_E^p|$ and of 3% on $|G_M^n|$ and of 10% on $|G_E^n|$ can be achieved from threshold up to $E_{BEAM} \approx 1.2$ GeV with an integrated luminosity of 100 $pb^{-1}$ per beam energy.

The projected values (with the relative errors) of the electromagnetic proton FFs and of the neutron magnetic FF compared with presently available data are shown in Fig. 6 and 7, respectively. As it can see, the proposed experiment will provide the first proton electric FF measurement and data for the proton magnetic FF with higher precision compared to the available measurements (especially close to the threshold). As concern the neutron, also in this case the proposed experiment will provide data on magnetic FF with high accuracy in a region in which they are scarce and with a big error.

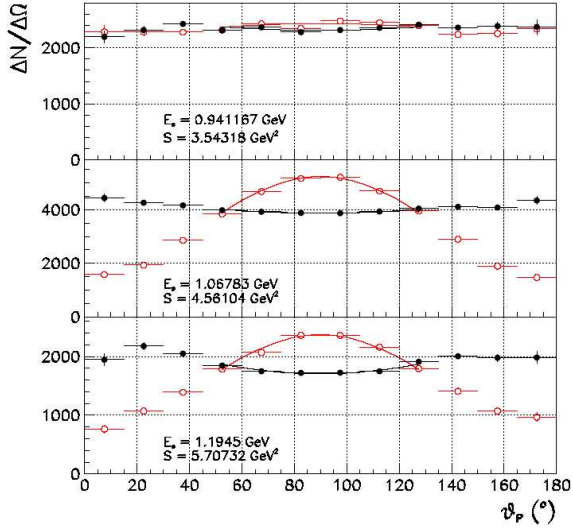
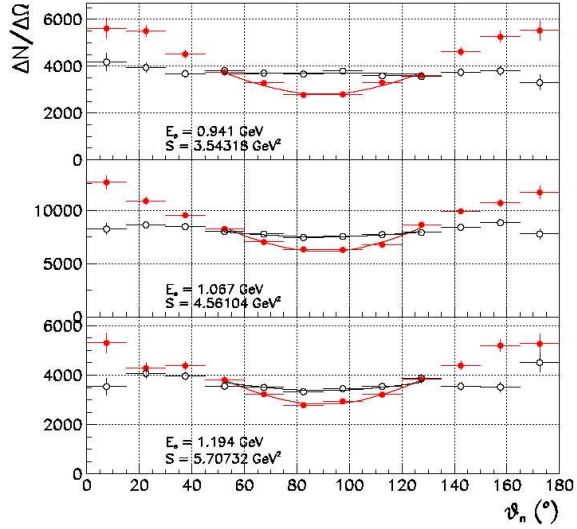

Figure 4. Projected angular distributions for $e^+e^- \to p\bar{p}$ at three different beam energies, for a detection efficiency $\epsilon_p = 0.85$ and an integrated luminosity of $100\,pb^{-1}$ per beam energy and assuming that either the electric and magnetic FFs are equal (full circles) or the ratio $|G_E^p|/|G_M^p|$ is given by the dispersion relations of ref. [3] (empty circles). The full line is a fit of the data with eq. (7) in the FINUDA angular range.

Figure 5. Projected angular distribution for $e^+e^- \to n\bar{n}$ at three different beam energies, assuming the electric and magnetic FFs are equal (full circles) or $|G_E^n| = 0$ (open circles), a detection efficiency $\epsilon_p = 0.15$ and an integrated luminosity of $100pb^{-1}$ per beam energy. The full line is a fit of the data to eq. (7) over the FINUDA angular coverage.

In conclusion, with an integrated luminosity of $3.5 fb^{-1}$ and using the available FINUDA detector with a small modifications and the inclusion of a polarimeter it is possible to obtain the following results:

- The first accurate measurement of the proton time-like form factors $|G_E^p|$ and $|G_M^p|$;

- The first measurement of the outgoing proton polarization, to get the relative phase between $|G_E^p|$ and $|G_M^p|$;

- The first measurement of the two photon contribution from the proton angular distributions asymmetry;

- The first measurement of the neutron time-like form factors $|G_E^n|$ and $|G_M^n|$;

- The first measurement of the outgoing neutron polarization, to get the relative phase between $|G_E^p|$ and $|G_M^p|$;

- The first complete measurement (moduli and phase) of the time-like strange baryon form factors.

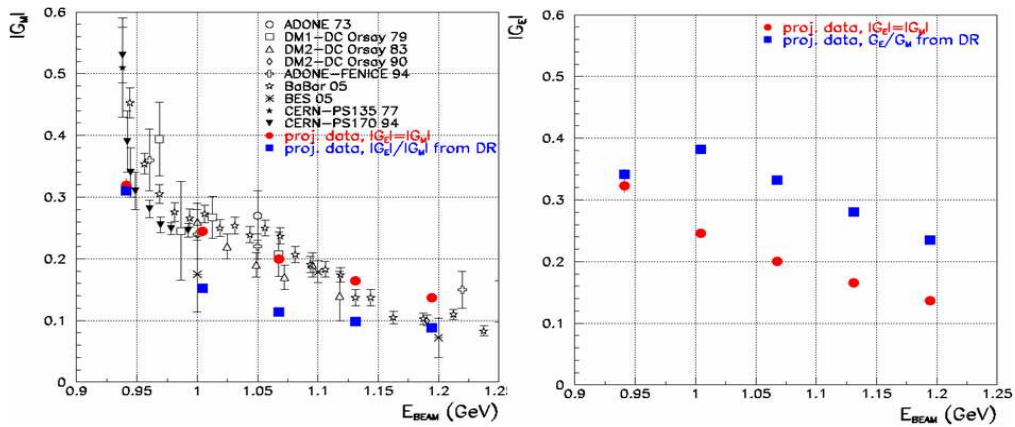

Figure 6. *Projected proton magnetic and electric FFs for the two hypothesis on the electric to magnetic FFs ratio ($|G_E^p| = |G_M^p|$: full circles and $|G_E^p|/|G_M^p|$ from dispersion relation [3]: full squares), compared with actual experimental values.*

The possibility to use KLOE (the other detector hosted in the DAΦNE storage ring) for this proposal has not been explored so far (it requires major modification for this measurement) but will be considered.

## 5. CONCLUSION

The internal structure of the nucleon is a defining problem for nuclear physics. One of the most basic observables which reflect the composite nature of the nucleon are its electromagnetic form factors which characterize its charge and magnetization distributions. In spite of being under investigation by more than fifty years, nucleon form factors are far from being fully exploited and recent results have shown a dramatically different picture of their knowledge.

In the time-like region form factors are poorly known experimentally and their extraction are based on specific assumptions for the electric to magnetic form factor ratio. The neutron data are very scarce and polarization data, directly connected with the complex phases of form factors, are not available at all.

In view of the possible increase of the energy of the DAΦNE collider at Frascati up to above the nucleon-antinucleon threshold, a physics program to perform the full determination of nucleon and hyperon form factors in the time sector has been proposed.

A preliminary counting rate estimate with the FINUDA detector has been presented. This detector is well suitable for the measurement, without major modifications and fulfills in an easy and satisfactory way all the main requirements of the proposed program.

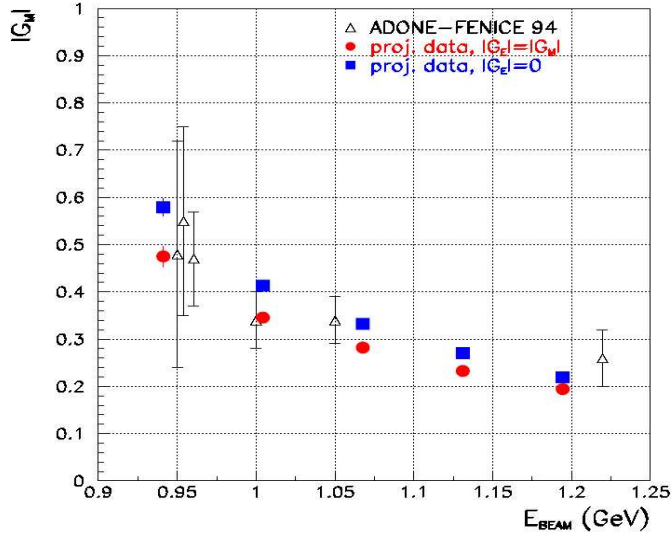

Figure 7. *Projected neutron magnetic FF results for the two hypothesis on the electric to magnetic FFs ratio ($|G_E^n| = |G_M^n|$: full circles and $|G_E^n| = 0$: full squares), compared with FENICE experimental data [18].*

An integrated luminosity of the order of $3.5 fb^{-1}$ should allow the measurement of $|G_M|$ and $|G_E|$ at the few percent level for the proton and below 10% for the neutron and polarization measurements will allow the first determination of the relative phase between the electric and magnetic form factors.